%% file: main.tex
\documentclass[
]{ceurart}

\usepackage{listings}
\usepackage{tabularx}
\usepackage{xurl}

\begin{document}

\copyrightyear{2024}
\copyrightclause{Copyright for this paper by its authors.
  Use permitted under Creative Commons License Attribution 4.0
  International (CC BY 4.0).}

\conference{EAmSI24: Edge AI meets swarm intelligence,
  September 18, 2024, Dubrovnik, Croatia}

\title{Experimental comparison of graph-based approximate nearest neighbor search algorithms on edge devices}

\author[1]{Ali Ganbarov}[%
orcid=0009-0000-7543-2391,
email=ali.ganbarov@tu-berlin.de,
]

\author[1]{Jicheng Yuan}[%
orcid=0009-0002-4448-2809,
email=jicheng.yuan@tu-berlin.de,
]

\author[1]{Anh Le-Tuan}[%
orcid=0000-0003-2458-607X,
email=anh.letuan@tu-berlin.de,
]

\author[1,2]{Manfred Hauswirth}[%
orcid=0000-0002-1839-0372,
email=manfred.hauswirth@tu-berlin.de,
]

\author[1,2]{Danh Le-Phuoc}[%
orcid=0000-0003-2480-9261,
email=danh.lephuoc@tu-berlin.de,
]

\address[1]{Open Distributed Systems, Technical University of Berlin}
\address[2]{Fraunhofer Institute for Open Communication Systems, Berlin, Germany}

\begin{abstract}
  In this paper, we present an experimental comparison of various graph-based approximate nearest neighbor (ANN) search algorithms deployed on edge devices for real-time nearest neighbor search applications, such as smart city infrastructure and autonomous vehicles. To the best of our knowledge, this specific comparative analysis has not been previously conducted. While existing research has explored graph-based ANN algorithms, it has often been limited to single-threaded implementations on standard commodity hardware. Our study leverages the full computational and storage capabilities of edge devices, incorporating additional metrics such as insertion and deletion latency of new vectors and power consumption. This comprehensive evaluation aims to provide valuable insights into the performance and suitability of these algorithms for edge-based real-time tracking systems enhanced by nearest-neighbor search algorithms.
\end{abstract}


\maketitle

\section{Motivation and contribution}
\input{text/motivation}

\section{Background and Related Work}
\input{text/bg_and_related_work}

\section{Methodology}
\input{text/methodology}

\section{Experimental Results}
\input{text/setup}

\subsection{Results and Discussions}
\input{text/results}

\section{Conclusion and Future Work}
This study presents an experimental evaluation of various approximate nearest neighbor search (ANNS) algorithms across diverse hardware configurations, with a primary focus on edge devices. Our findings underscore several key insights: the performance superiority of specific algorithms tailored to particular devices and the suitability of different devices for distinct types of algorithms. Notably, our research demonstrates that cost-effective devices like the Raspberry Pi 4 can achieve competitive performance compared to more expensive and power-intensive Jetson devices in CPU-based tasks. Furthermore, we showcased the end-to-end workflow capabilities of DiskANN, particularly its support for true deletion of vectors. This feature proves invaluable for maintaining model relevance for potential shifts in data distribution over time and supports continuous on-device learning.

Looking ahead, our findings highlight several promising research directions. First, optimizing FAISS algorithms for resource-constrained edge devices, such as the Raspberry Pi 3 and Zero, is crucial. Enhancing these algorithms to function efficiently with limited memory and processing power could improve their practical use and reduce power consumption. Additionally, our study identifies performance differences among devices with varying CPU capabilities. Future research should investigate why some algorithms perform better on devices with lower CPU power and explore which hardware specifications most impact performance.

In conclusion, our study contributes to advancing the understanding of ANNS algorithm performance across edge devices, providing a foundation for optimizing these algorithms across diverse applications in machine learning and data-intensive tasks. 

\section{Acknowledgements}
This work is supported by the German Research Foundation (DFG) under the COSMO project (grant No. 453130567), and by the European Union’s Horizon WIDERA under the grant agreement No. 101079214 (AIoTwin),  by the Federal Ministry for Education and Research Germany under grant number 01IS18037A (BIFOLD), and RIA research and innovation program under the grant agreement No. 101092908 (SmartEdge).

\clearpage
\bibliography{main}

\end{document}

%% file: text/motivation.tex
Approximate Nearest Neighbor Search (ANNS) has become more crucial as the amount of data we have to handle keeps increasing rapidly. ANNS plays a key role in addressing the \(k\)-Nearest Neighbor Search (\(k\)-NNS) issue, where the task is to identify the \(k\) most similar vectors to a given query vector within a dataset. The easiest exact approach to \(k\)-NNS involves measuring distances between the query vector and all vectors in the dataset, followed by selecting the \(k\) vectors with the shortest distances. However, this approach is not feasible for large datasets due to the significant computational burden, which poses challenges for scalability. As datasets expand exponentially and the curse of dimensionality becomes a critical concern, precise NNS methods become inefficient and costly. Consequently, research has pivoted towards ANNS algorithms that substantially enhance efficiency while allowing for a slight decrease in accuracy. ANNS achieves a balance between speed and precision, minimizing computational requirements and facilitating scalability for extensive datasets. These algorithms utilize indexing to identify approximate nearest neighbors in high-dimensional spaces, providing a practical solution for numerous applications where exact results are not essential.



Nearest Neighbor Search (NNS) is fundamental in various sectors such as recommendation systems \cite{sarwar2001itembased}, data retrieval \cite{papadopoulos2006nearest}, information matching \cite{lulu2013images}. Although the brute force method is straightforward and intuitive, its computational cost increases significantly as data volume grows. ANNS approaches address this challenge by significantly reducing search times with only a minor sacrifice in accuracy, thereby balancing precision and latency to suit different application needs.



This paper presents an experimental evaluation of approximate nearest neighbor (ANNS) algorithms on edge devices for smart city applications, using data from street cameras provided by Conveqs and Aalto University in Helsinki. The captured data was preprocessed to extract object bounding boxes, which were then embedded into vectors for constructing an index graph for ANNS classification tasks. Our study reveals several key insights: the performance patterns of algorithms differ significantly between edge devices and servers, more powerful edge devices do not always yield more efficient algorithm execution, and cost-effective devices can achieve performance comparable to much more expensive counterparts.

This paper is structured as follows: Section 1 discusses the motivation and contributions of the study. Section 2 provides background information and related work done in the field. Section 3 outlines the methodology, including the metrics used and their justifications, the process for determining accuracy, the algorithm implementations, and the data utilized. Section 4 presents the experimental results and discussions. Finally, Section 5 concludes the paper and suggests directions for future work.

%% file: text/bg_and_related_work.tex
In this section, we will discuss Product Quantization for data compression to fit models into edge devices with smaller memory capacity and Approximate Nearest Neighbor Search algorithms for vector search to retrieve the closest items in the dataset to a given query. Additionally, we will give an overview of existing work.

\subsection{Background}
Product Quantization (PQ) is an effective technique for approximate nearest neighbor (ANN) search, ideal for edge devices \cite{jegou2010product}. It reduces storage and computational demands, making it suitable for resource-constrained environments. PQ splits high-dimensional vectors into lower-dimensional sub-vectors, which are independently quantized using codebooks, thus approximating the original data with reduced storage needs.

ANNS techniques are crucial in fields like data mining, AI, NLP \cite{xu2023nearest}, computer vision \cite{banerjee2019view}, information retrieval \cite{liu2007clustering}, and advertising \cite{li2023practice}. They are vital for clustering, classification, and dimensionality reduction. ANNS algorithms are classified into hashing-based \cite{terasawa2007spherical}, tree-based \cite{arora2018hdindex}, quantization-based \cite{zhang2019grip}, and graph-based methods \cite{fu2019fast}. Graph-based methods, in particular, offer high recall rates and reduced search times, making them valuable for many practical applications.

\begin{itemize}
\item \textbf{Hierarchical Navigable Small World (HNSW)}
\cite{malkov2018efficient} is an approximate nearest neighbor algorithm that builds a multi-layer graph. It organizes data into hierarchical layers, with each layer containing fewer points. The algorithm uses a greedy search from the top layer down, efficiently narrowing the search space. This structure allows HNSW to achieve high recall rates and fast search times, making it ideal for large-scale, high-dimensional data retrieval.

\item \textbf{Vamana}
\cite{subramanya2019diskann} is an approximate nearest neighbor algorithm that builds a navigable small-world graph. Unlike HNSW, Vamana focuses on balanced connectivity, ensuring each node links to both close and distant neighbors. This approach enables efficient graph traversal and rapid identification of nearest neighbors, making it well-suited for large-scale, high-dimensional datasets while keeping computational costs low.

\item \textbf{Inverted-file search (IVF)} 
\cite{coster2002inverted} is a two-stage search algorithm that combines a coarse quantizer with Product Quantization (PQ). First, IVF partitions the database into \(k_0\) Voronoi cells and selects the \(w\) nearest cells for a query vector. In the second stage, PQ is applied only to vectors in these selected cells, which reduces the search space and computational costs, balancing efficiency and accuracy.

\item \textbf{Locality-Sensitive Hashing (LSH)} 
\cite{dasgupta2011fast} improves approximate nearest neighbor search by mapping high-dimensional data into low-dimensional buckets using hash functions. Similar points are likely to fall into the same bucket, reducing the search space. LSH computes hash values for the query vector and retrieves candidates from the relevant buckets, thus enhancing efficiency and minimizing computational costs. This method balances search accuracy and speed, making it suitable for large-scale and real-time applications.

 \end{itemize}

\subsection{Related Work}
In a comprehensive study, \cite{wang2021comprehensive} provides a detailed comparison of graph-based approximate nearest neighbor search (ANNS) algorithms. The authors implemented these algorithms from scratch in C++ and evaluated them theoretically without using parallel programming or GPU techniques, which are essential for practical optimization. The study compares 13 algorithms across various datasets, offering insights into their performance, strengths, and weaknesses.

\cite{jiang2018survey} assesses graph-based ANNS algorithms with quantization techniques for real-time streaming data in fog computing. Their focus is on algorithmic redesign for FPGA architectures to reduce network congestion. In contrast, our research targets the deployment of ANNS algorithms on edge devices, utilizing CPU and GPU resources. We evaluate algorithms in their native state, contrasting with FPGA-focused modifications in the referenced study \cite{jiang2018survey}.

%% file: text/methodology.tex
In this section, we will discuss the crucial metrics that are employed in real-time performance-critical scenarios such as smart city applications. Real-time object detection systems, essential for smart traffic management, public safety, and environmental monitoring, demand efficient algorithms to be able to process incoming data swiftly. Our study incorporates state-of-the-art ANNS algorithms to process large amounts of data collected from street cameras and evaluates them based on metrics aimed at optimizing real-time performance. The following subsections will discuss the accuracy measurement technique, evaluation metrics, algorithm implementations, and data.

\subsection{Accuracy Measure}
To evaluate the accuracy of graph-based nearest neighbor search algorithms, we employ a process where the \( K \) nearest neighbors for each query are retrieved, and their class labels are analyzed. Given a query point \( q \), the algorithm identifies its \( K \) nearest neighbors in the dataset
\begin{math}
\{x_1, x_2, \ldots, x_N\} 
\end{math}
. Let 
\begin{math}
 \mathcal{N}_K(q)
\end{math}
denote the set of these \( K \) nearest neighbors. The class labels of these neighbors are then counted to determine the most frequent class label, which is assigned as the predicted class label for \( q \). Formally, let \( y_i \) represent the class label of the \( i \)-th nearest neighbor. The predicted class label \( \hat{y}(q) \) is given by:

\begin{displaymath}
\hat{y}(q) = \underset{c \in \mathcal{C}}{\arg\max} \sum_{x_i \in \mathcal{N}_K(q)} \mathbb{I}(y_i = c)
\end{displaymath}

where  
\begin{math}
\mathcal{C} 
\end{math}
is the set of all possible class labels and \( \mathbb{I}(\cdot) \) is the indicator function, which equals 1 if the argument is true and 0 otherwise. The overall accuracy \( A \) of the algorithm is calculated as the proportion of correctly classified query points out of the total number of query points \( Q \)

\begin{displaymath}
A = \frac{1}{Q} \sum_{j=1}^{Q} \mathbb{I}(\hat{y}(q_j) = y(q_j))
\end{displaymath}

Here, \( y(q_j) \) is the true class label of the \( j \)-th query point \( q_j \). This method provides a measure of how well the graph-based nearest neighbor search algorithm can classify points based on the labels of their nearest neighbors.

\subsection{Real-time Performance Metrics}
In evaluating graph-based approximate nearest neighbor search (ANNS) algorithms for real-time object detection on edge devices in smart city applications, the performance metrics chosen are Queries per Second (QPS), insertion speed, deletion speed, and power consumption. These metrics are essential for assessing the algorithms' effectiveness and efficiency in practical scenarios.

\begin{itemize}
\item \textbf{Queries per Second (QPS)} 
measures how many search queries the algorithm can handle per second. High QPS is crucial for real-time object detection, ensuring timely responses in dynamic environments such as traffic monitoring and public safety.

\item \textbf{Insertion Speed}
refers to the time taken to add new data points into the ANNS structure. Efficient insertion is important for dynamic datasets, like real-time video feeds and sensor data, allowing the system to adapt quickly to new information.

\item \textbf{Deletion Speed} 
measures how quickly outdated or irrelevant data points can be removed. Fast deletion helps manage storage and keeps the system responsive by preventing data buildup, which is crucial in environments with limited resources.

\item \textbf{Power Consumption} 
evaluates the energy efficiency of the algorithm. Low power consumption is important for edge devices in resource-constrained settings to reduce operational costs and extend device lifespan.
\end{itemize}

The selected performance metrics - queries per second, insertion speed, deletion speed, and power consumption - provide a comprehensive evaluation framework for assessing the suitability of graph-based ANNS algorithms in real-time object detection for smart city applications. These metrics address both the computational performance and practical deployment considerations, ensuring that the chosen algorithms can deliver accurate, timely, and energy-efficient solutions in dynamic urban environments.

\subsection{Algorithm Implementation}
FAISS (Facebook AI Similarity Search) \cite{johnson2019billion} and DiskANN \cite{subramanya2019diskann} are key tools for efficient approximate nearest neighbor search (ANNS). FAISS, from Facebook AI Research, offers various indexing methods optimized for CPU and GPU, making it suitable for real-time applications and large datasets. DiskANN focuses on disk-based indexing, using methods like Vamana to efficiently organize and query large datasets with limited memory. DiskANN's approach allows for fast searches by constructing a structured graph representation of data on disk, ideal for memory-constrained environments. Both FAISS and DiskANN provide effective solutions for diverse ANNS needs, supporting both research and industrial applications.

\subsection{Data}
In this study, we collected data from the test platform set up by Conveqs and Aalto University in Helsinki, near Länsisatama (Western Harbor), Jätkäsaari~\footnote{\url{https://www.smart-edge.eu/wp-content/uploads/2023/10/SmartEdge-D2.1-First-definition-or-requirements-architecture-and-use-cases.pdf}}. This platform consists of 17 roadside cameras along with radars. For this experiment, four cameras (No. $269_2$, $269_3$, $270_2$, $270_3$) were selected based on their strategic placement, capturing diverse traffic patterns and environmental conditions. These cameras were positioned at high-traffic intersections to ensure a comprehensive dataset covering various vehicle types and pedestrian activities. Due to the large volume of data, we extracted one frame for every ten frames. In total, we have 12,451 frames across ten classes. The box predictions are from the SOTA 2D detection model Co-DETR~\cite{zong2023detrs} (pre-trained on MS-COCO~\cite{lin2014microsoft}), and we filtered the predictions from the classification head with a score greater than 0.5 as our pseudo-label. This method resulted in a total of 161,805 valid bounding boxes, which we split into 85\% for training and 15\% for evaluation.

To ensure the quality and consistency of the dataset, we established a preprocessing pipeline. This included steps such as removing corrupted frames, correcting for lens distortion, and normalizing image dimensions to $224\times224$ pixels to match the input requirements of the classification models which were pre-trained on ImageNet~\cite{deng2009imagenet}. In addition, data augmentation such as random cropping, horizontal flipping, and brightness adjustments were applied to increase the variability and robustness of the dataset.

For embedding generation, we employed the ResNet50~\cite{he2016deep}, a convolutional neural network pre-trained on the ImageNet dataset, as our feature extractor due to its proven performance in various image recognition tasks and its effective performance in capturing hierarchical features through residual learning. To serve as a feature extractor, we removed the final fully connected layer and replaced it with a global average pooling layer, followed by a dense layer producing a 2048-dimensional embedding vector for each image. This configuration ensures that the generated embeddings encapsulate the consistent features of the images.

%% file: text/setup.tex
\subsection{Setup}
In our experimental setup to compare the performance of graph-based nearest-neighbor search algorithms, we utilized several edge devices, including four from the NVIDIA Jetson family: Nano, TX2, Xavier AGX, and Orin. Additionally, we incorporated other edge devices such as the Raspberry Pi 4, Raspberry Pi 3, and Raspberry Pi Zero. Tables \ref{tab:jetson} and \ref{tab:raspberrypi} provide a comparative analysis of these devices and their capabilities. As observed, devices with higher computational power, such as the Jetson Xavier AGX and Jetson Orin, exhibit increased power consumption, underscoring the trade-off between performance and energy efficiency.

These devices are widely used in edge AI application processing scenarios. For instance, in graph-based approximate nearest neighbor search tasks for embedded images with bounding boxes, the computational demands necessitate efficient processing power balanced with manageable energy consumption. This is particularly relevant in applications such as real-time object detection and tracking in surveillance systems, where quick and accurate identification of objects is crucial.

\begin{table}[ht!]
  \caption{Comparison of Jetson Family Devices}
  \label{tab:jetson}
  \begin{tabular}{cccccl}
    \toprule
    Jetson&GPU&Clock Speed&Memory&TOPS&Power \\
    \midrule
    Nano & 128-core Maxwell & 1.43 GHz & 4 GB & 0.5 & 5-10W \\
    TX2 & 256-core Pascal & 2 GHz & 8 GB & 1.3 & ~7.5W \\
    Xavier & 512-core Volta + 64 Tensor Cores & 2.26 GHz & 32 GB & 32 & 30W \\
    Orin & 2048-core Ampere + 64 Tensor Cores & 2.0 GHz & 64 GB & 200 & 30-40W \\
  \bottomrule
\end{tabular}
\end{table}

In autonomous vehicles, Jetson devices like the Jetson Xavier AGX and Orin process data from sensors such as cameras and LIDAR to identify and track objects, navigate environments, and make real-time decisions. Their high-performance GPUs and AI capabilities enable sophisticated algorithms for tasks such as object detection and lane recognition in advanced driver assistance systems (ADAS). Utilizing a single Jetson AGX Xavier board integrates computing resources—CPUs, GPUs, and DLA cores—reducing costs, power consumption, and cross-device communication issues, ensuring effective ADAS operation \cite{nguyen2024optimizing}.

In healthcare, Jetson devices facilitate portable medical imaging and diagnostics \cite{pace2018edge}. They enable real-time processing of ultrasound or MRI images on-site, reducing reliance on centralized servers. The Raspberry Pi 3, noted for its performance and cost-effectiveness, is also used in healthcare for flexible, robust solutions that minimize data transmission and processing time.

\begin{table}[ht!]
\caption{Comparison of Raspberry Pi Family Devices}
  \label{tab:raspberrypi}
  \begin{tabular}{cccl}
    \toprule
    Raspberry&Clock Speed&Memory&Power\\
    \midrule
    Pi 4 & 1.5 GHz & 8 GB & 3.4W \\
    Pi 3 & 1.2 GHz & 1 GB & 3.7W \\
    Pi Zero & 1 GHz & 512 MB & 1.1W \\
  \bottomrule
\end{tabular}
\end{table}

For smart city applications, Raspberry Pi devices can be effectively utilized in video surveillance systems. Their low power consumption and cost-effectiveness make them ideal for extensive deployment in both public and private areas, enabling real-time data collection and analysis. By integrating Raspberry Pi into these systems, the devices can perform complex recognition tasks independently, thereby reducing the need for centralized server resources \cite{kavalionak2019distributed}.

These examples illustrate the versatility and importance of selecting the appropriate edge computing hardware for various AI and machine learning applications, balancing computational power with energy efficiency to meet the specific needs of different use cases.


%% file: text/results.tex
In our experimental setup, we utilized FAISS implementations of IVF, LSH, HNSW, and PQ algorithms. For Vamana, DiskANN was used. Construction and parameter optimization were conducted on a server equipped with 28 CPU cores and 2TB of RAM. PQ was applied across all algorithms to ensure compatibility with memory constraints of edge devices. However, integrating PQ with DiskANN using standard tools posed challenges, and we were unable to deploy it on edge devices. Therefore, we will first present comparative ANNS performance results on our server and subsequently evaluate FAISS algorithms on edge devices. Table \ref{tab:construction} provides details on the build timew and index size of the most accurate configurations achieved on the server. Following construction, these indices were stored and transferred to edge devices for conducting nearest-neighbor searches on the test dataset.

\begin{table}[ht!]
  \caption{Construction time and index size of $\sim$1M vectors}
  \label{tab:construction}
  \begin{tabular}{ccl}
    \toprule
    Algorithm&Build time (s)&Index size (MB)\\
    \midrule
    PQ & 49.7 & 18 \\
    LSH & 0.9 & 33 \\
    HNSW-PQ & 290 & 25 \\
    IVF-PQ & 40 & 26 \\
    DiskANN & 455 & 8,000 \\
  \bottomrule
\end{tabular}
\end{table}

\begin{figure}
  \centering
  \includegraphics[width=100mm]{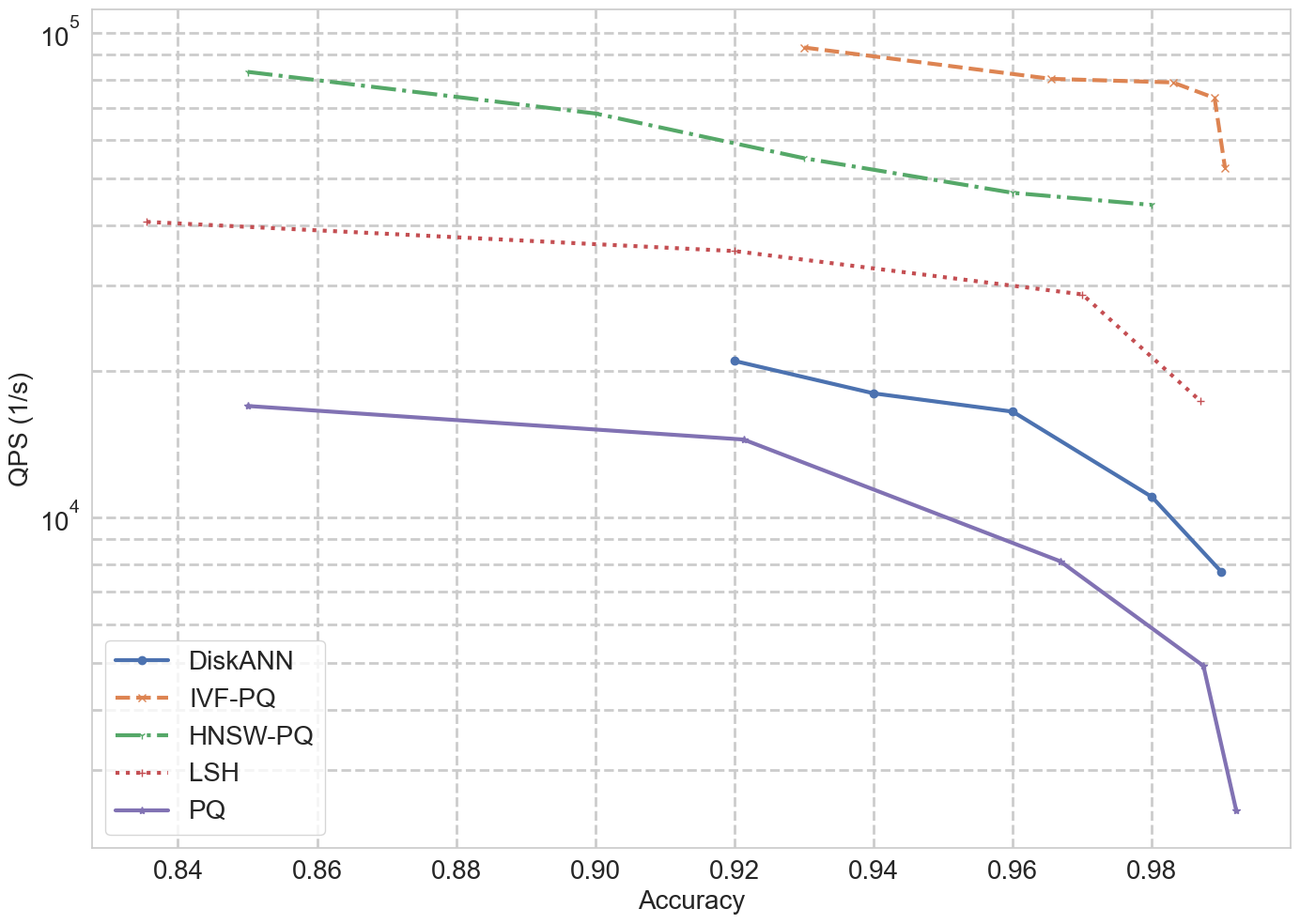}
  \caption{Server comparison using CPU}
  \label{server-comparison}
\end{figure}

Figure \ref{server-comparison} illustrates a comparison of these algorithms on a server running on CPU, depicting the tradeoff between Queries Per Second (QPS) and accuracy. The graph highlights the balance between QPS and accuracy for each algorithm. Notably, DiskANN, which does not use any data compression and operates on an index built from the original data vectors, shows competitive performance relative to other approximate nearest neighbor search (ANNS) methods that utilize Product Quantization (PQ) for data compression. Furthermore, it is evident that FAISS algorithms, even with compression, can achieve high accuracy while maintaining high QPS.

\begin{figure}
  \centering
  \includegraphics[width=\linewidth]{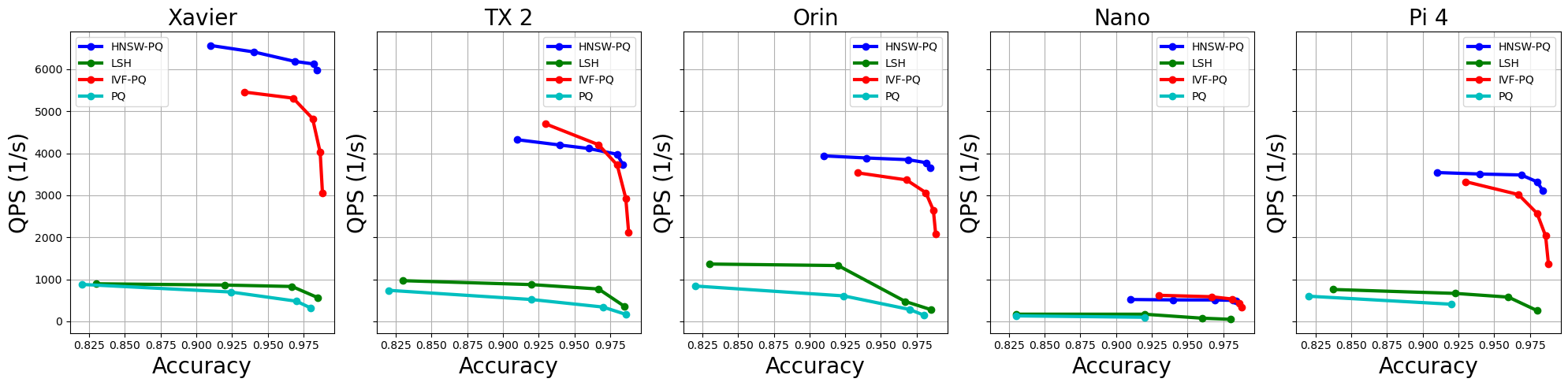}
  \caption{Edge device comparison using CPU}
  \label{device-comparison}
\end{figure}

\begin{figure}
  \centering
  \includegraphics[width=\linewidth]{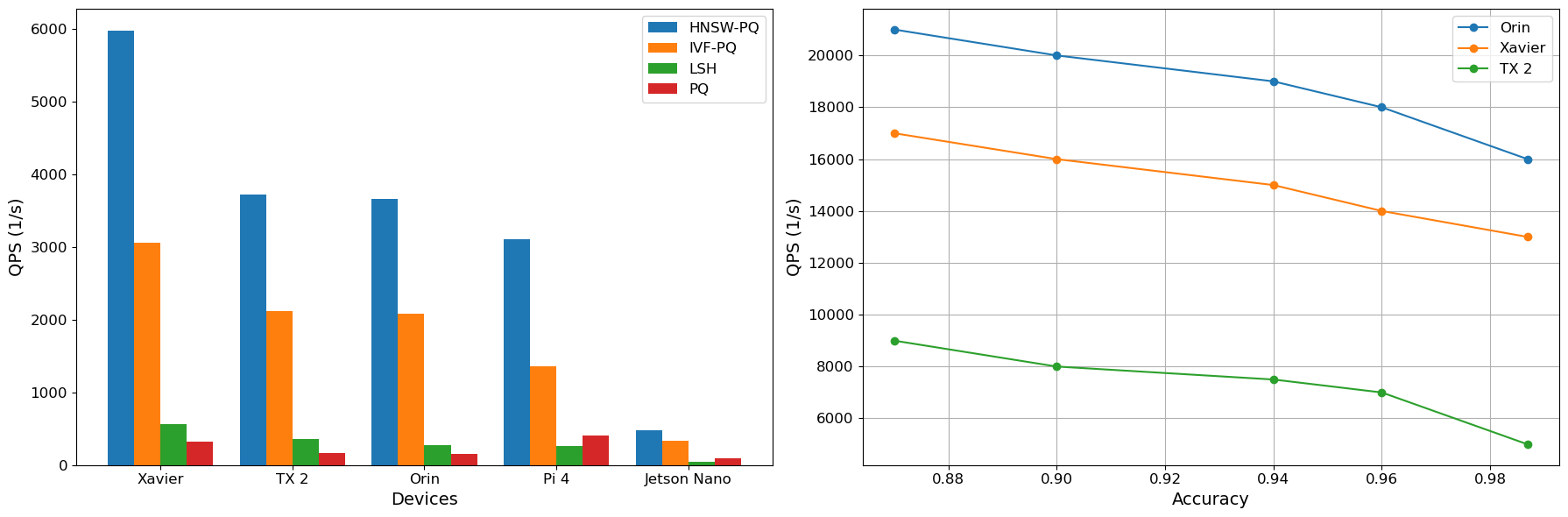}
  \caption{CPU(left) and GPU(right) performance comparison across devices}
  \label{qps-comparison}
\end{figure}

Figure \ref{device-comparison} shows the performance of various algorithms on edge devices using CPU. FAISS was not supported on the Raspberry Pi 3 and Pi Zero, but the Raspberry Pi 4 proved to be a strong competitor in CPU-based approximate nearest-neighbor search. Despite similar CPU clock speeds and double the memory of the Pi 4, it performed over three times better than the Jetson Nano while being about one-fourth the price. The Jetson Xavier also outperformed the Jetson Orin by 1.5 times in CPU utilization for HNSW-PQ, despite having less memory and fewer CPU cores (8 versus 12). This discrepancy may be due to Xavier's larger L2 cache (8MB vs. Orin's 3MB). On the Jetson TX2, the IVF-PQ algorithm achieved higher QPS than HNSW-PQ at lower accuracy levels. Generally, HNSW-PQ outperformed IVF-PQ across all devices, unlike the server results. Figure \ref{qps-comparison} (left) shows that the Jetson Xavier had the highest throughput for both HNSW-PQ and IVF-PQ. The Raspberry Pi 4's CPU performance is comparable to the Jetson TX2 and Orin. Additionally, PQ performed better than LSH on lower-power devices like the Raspberry Pi 4 and Jetson Nano.

For GPU comparison, we utilized the HNSW-PQ algorithm implementation of ANNS with the FAISS library to conduct inference on GPUs. The outcomes are illustrated in Figure \ref{qps-comparison} (right). As anticipated, the Queries Per Second (QPS) metric demonstrated higher performance on more powerful devices. It is intriguing that GPU acceleration provided a threefold improvement in throughput compared to CPU implementations. However, it's important to consider that GPU-based inference consumes double the power of CPU-based inference on these devices.

\begin{figure}
  \centering
  \includegraphics[width=100mm]{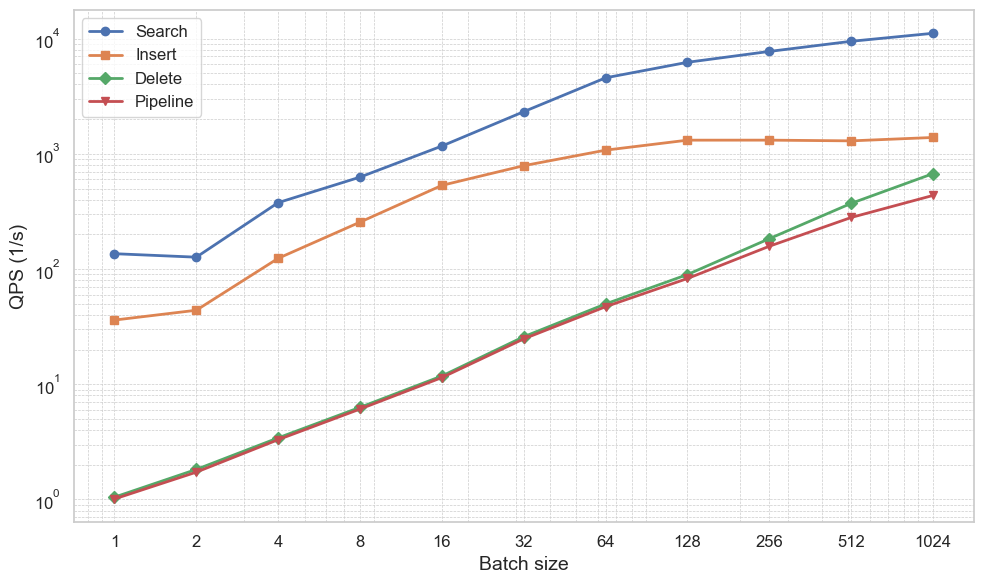}
  \caption{DiskANN pipeline performance}
  \label{diskann-pipeline}
\end{figure}

DiskANN's lack of support for compression prevented us from fitting its index onto edge devices, so we evaluated it on our server. DiskANN allows true vector deletion, unlike FAISS, which uses "soft" deletion—vectors are marked as deleted but remain in the index structure. We developed a test pipeline for insertion and deletion with DiskANN, simulating real-world data scenarios. Figure \ref{diskann-pipeline} shows QPS results for search, insertions, deletions, and overall performance across various batch sizes. Deletion throughput significantly affects pipeline performance, similar to insertion operations, due to the costly re-indexing process required. Larger batch sizes generally improved throughput, with search and insertion operations effectively managing batches up to 128 before plateauing. The diminishing returns in search and insertion had a greater impact compared to the slowdown in deletions.

%% file: main.bbl
\begin{thebibliography}{26}
\expandafter\ifx\csname natexlab\endcsname\relax\def\natexlab#1{#1}\fi
\providecommand{\url}[1]{\texttt{#1}}
\providecommand{\href}[2]{#2}
\providecommand{\path}[1]{#1}
\providecommand{\DOIprefix}{doi:}
\providecommand{\ArXivprefix}{arXiv:}
\providecommand{\URLprefix}{URL: }
\providecommand{\Pubmedprefix}{pmid:}
\providecommand{\doi}[1]{\href{http://dx.doi.org/#1}{\path{#1}}}
\providecommand{\Pubmed}[1]{\href{pmid:#1}{\path{#1}}}
\providecommand{\bibinfo}[2]{#2}
\ifx\xfnm\relax \def\xfnm[#1]{\unskip,\space#1}\fi
\bibitem[{Sarwar et~al.(2001)Sarwar, Karypis, Konstan, and Riedl}]{sarwar2001itembased}
\bibinfo{author}{B.~Sarwar}, \bibinfo{author}{G.~Karypis}, \bibinfo{author}{J.~Konstan}, \bibinfo{author}{J.~Riedl},
\newblock \bibinfo{title}{Itembased collaborative filtering recommendation algorithms},
\newblock in: \bibinfo{booktitle}{Proceedings of the 10th International Conference on World Wide Web}, \bibinfo{year}{2001}, pp. \bibinfo{pages}{285--295}.
\bibitem[{Papadopoulos and Manolopoulos(2006)}]{papadopoulos2006nearest}
\bibinfo{author}{A.~N. Papadopoulos}, \bibinfo{author}{Y.~Manolopoulos}, \bibinfo{title}{Nearest Neighbor Search: A Database Perspective}, \bibinfo{publisher}{Springer Science \& Business Media}, \bibinfo{year}{2006}.
\bibitem[{Lulu et~al.(2013)Lulu, Guohua, Kang, and Ajing}]{lulu2013images}
\bibinfo{author}{Z.~Lulu}, \bibinfo{author}{G.~Guohua}, \bibinfo{author}{L.~Kang}, \bibinfo{author}{H.~Ajing},
\newblock \bibinfo{title}{Images matching algorithm based on surf and fast approximate nearest neighbor search},
\newblock \bibinfo{journal}{Application Research of Computers} \bibinfo{volume}{30} (\bibinfo{year}{2013}) \bibinfo{pages}{921--923}.
\bibitem[{Jegou et~al.(2010)Jegou, Douze, and Schmid}]{jegou2010product}
\bibinfo{author}{H.~Jegou}, \bibinfo{author}{M.~Douze}, \bibinfo{author}{C.~Schmid},
\newblock \bibinfo{title}{Product quantization for nearest neighbor search},
\newblock \bibinfo{journal}{IEEE transactions on pattern analysis and machine intelligence} \bibinfo{volume}{33} (\bibinfo{year}{2010}) \bibinfo{pages}{117--128}.
\bibitem[{Xu et~al.(2023)Xu, Alon, and Neubig}]{xu2023nearest}
\bibinfo{author}{F.~F. Xu}, \bibinfo{author}{U.~Alon}, \bibinfo{author}{G.~Neubig},
\newblock \bibinfo{title}{Why do nearest neighbor language models work?},
\newblock \bibinfo{journal}{arXiv preprint arXiv:2301.02828}  (\bibinfo{year}{2023}). \bibinfo{note}{ArXiv:2301.02828 [cs]}.
\bibitem[{Banerjee et~al.(2019)Banerjee, Connolly, Lisin, Briggs, and Munich}]{banerjee2019view}
\bibinfo{author}{N.~Banerjee}, \bibinfo{author}{R.~C. Connolly}, \bibinfo{author}{D.~Lisin}, \bibinfo{author}{J.~Briggs}, \bibinfo{author}{M.~E. Munich},
\newblock \bibinfo{title}{View management for lifelong visual maps},
\newblock in: \bibinfo{booktitle}{2019 IEEE/RSJ International Conference on Intelligent Robots and Systems (IROS)}, \bibinfo{publisher}{IEEE}, \bibinfo{year}{2019}, pp. \bibinfo{pages}{7871--7878}.
\bibitem[{Liu et~al.(2007)Liu, Rosenberg, and Rowley}]{liu2007clustering}
\bibinfo{author}{T.~Liu}, \bibinfo{author}{C.~Rosenberg}, \bibinfo{author}{H.~A. Rowley},
\newblock \bibinfo{title}{Clustering billions of images with large scale nearest neighbor search},
\newblock in: \bibinfo{booktitle}{2007 IEEE Workshop on Applications of Computer Vision (WACV '07)}, \bibinfo{publisher}{IEEE}, \bibinfo{year}{2007}, pp. \bibinfo{pages}{28--28}.
\bibitem[{Li et~al.(2023)Li, Zhao, Wang, Xia, Wu, and Peng}]{li2023practice}
\bibinfo{author}{P.~Li}, \bibinfo{author}{W.~Zhao}, \bibinfo{author}{C.~Wang}, \bibinfo{author}{Q.~Xia}, \bibinfo{author}{A.~Wu}, \bibinfo{author}{L.~Peng},
\newblock \bibinfo{title}{Practice with graph-based ann algorithms on sparse data: Chi-square two-tower model, hnsw, sign cauchy projections},
\newblock \bibinfo{journal}{arXiv preprint arXiv:2306.07607}  (\bibinfo{year}{2023}). \bibinfo{note}{ArXiv:2306.07607 [cs, stat]}.
\bibitem[{Terasawa and Tanaka(2007)}]{terasawa2007spherical}
\bibinfo{author}{K.~Terasawa}, \bibinfo{author}{Y.~Tanaka},
\newblock \bibinfo{title}{Spherical lsh for approximate nearest neighbor search on unit hypersphere},
\newblock in: \bibinfo{booktitle}{Workshop on Algorithms and Data Structures}, \bibinfo{publisher}{Springer}, \bibinfo{year}{2007}, pp. \bibinfo{pages}{27--38}.
\bibitem[{Arora et~al.(2018)Arora, Sinha, Kumar, and Bhattacharya}]{arora2018hdindex}
\bibinfo{author}{A.~Arora}, \bibinfo{author}{S.~Sinha}, \bibinfo{author}{P.~Kumar}, \bibinfo{author}{A.~Bhattacharya},
\newblock \bibinfo{title}{Hd-index: Pushing the scalability-accuracy boundary for approximate knn search in high-dimensional spaces},
\newblock \bibinfo{journal}{arXiv preprint arXiv:1804.06829}  (\bibinfo{year}{2018}).
\bibitem[{Zhang and He(2019)}]{zhang2019grip}
\bibinfo{author}{M.~Zhang}, \bibinfo{author}{Y.~He},
\newblock \bibinfo{title}{Grip: Multi-store capacity-optimized high-performance nearest neighbor search for vector search engine},
\newblock in: \bibinfo{booktitle}{Proceedings of the 28th ACM International Conference on Information and Knowledge Management}, \bibinfo{year}{2019}, pp. \bibinfo{pages}{1673--1682}.
\bibitem[{Fu et~al.(2019)Fu, Xiang, Wang, and Cai}]{fu2019fast}
\bibinfo{author}{C.~Fu}, \bibinfo{author}{C.~Xiang}, \bibinfo{author}{C.~Wang}, \bibinfo{author}{D.~Cai},
\newblock \bibinfo{title}{Fast approximate nearest neighbor search with the navigating spreading-out graph},
\newblock \bibinfo{journal}{Proceedings of the VLDB Endowment} \bibinfo{volume}{12} (\bibinfo{year}{2019}) \bibinfo{pages}{461--474}.
\bibitem[{Malkov and Yashunin(2018)}]{malkov2018efficient}
\bibinfo{author}{Y.~A. Malkov}, \bibinfo{author}{D.~A. Yashunin},
\newblock \bibinfo{title}{Efficient and robust approximate nearest neighbor search using hierarchical navigable small world graphs},
\newblock \bibinfo{journal}{IEEE Transactions on Pattern Analysis and Machine Intelligence} \bibinfo{volume}{42} (\bibinfo{year}{2018}) \bibinfo{pages}{824--836}.
\bibitem[{Subramanya et~al.(2019)Subramanya, Suhas, and et~al.}]{subramanya2019diskann}
\bibinfo{author}{J.~Subramanya}, \bibinfo{author}{Suhas}, \bibinfo{author}{et~al.},
\newblock \bibinfo{title}{Diskann: Fast accurate billion-point nearest neighbor search on a single node},
\newblock in: \bibinfo{booktitle}{Advances in Neural Information Processing Systems 32}, \bibinfo{year}{2019}.
\bibitem[{C{\"o}ster and Svensson(2002)}]{coster2002inverted}
\bibinfo{author}{R.~C{\"o}ster}, \bibinfo{author}{M.~Svensson},
\newblock \bibinfo{title}{Inverted file search algorithms for collaborative filtering},
\newblock in: \bibinfo{booktitle}{Proceedings of the 25th annual international ACM SIGIR conference on Research and development in information retrieval}, \bibinfo{year}{2002}, pp. \bibinfo{pages}{246--252}.
\bibitem[{Dasgupta et~al.(2011)Dasgupta, Kumar, and Sarl{\'o}s}]{dasgupta2011fast}
\bibinfo{author}{A.~Dasgupta}, \bibinfo{author}{R.~Kumar}, \bibinfo{author}{T.~Sarl{\'o}s},
\newblock \bibinfo{title}{Fast locality-sensitive hashing},
\newblock in: \bibinfo{booktitle}{Proceedings of the 17th ACM SIGKDD international conference on Knowledge discovery and data mining}, \bibinfo{year}{2011}, pp. \bibinfo{pages}{1073--1081}.
\bibitem[{Wang et~al.(2021)Wang, Xu, Yue, and Wang}]{wang2021comprehensive}
\bibinfo{author}{M.~Wang}, \bibinfo{author}{X.~Xu}, \bibinfo{author}{Q.~Yue}, \bibinfo{author}{Y.~Wang},
\newblock \bibinfo{title}{A comprehensive survey and experimental comparison of graph-based approximate nearest neighbor search},
\newblock \bibinfo{journal}{arXiv preprint arXiv:2101.12631}  (\bibinfo{year}{2021}).
\bibitem[{Jiang et~al.(2018)Jiang, Hu, Li, Yuan, Masood, Jelodar, Rabbani, and Wang}]{jiang2018survey}
\bibinfo{author}{X.~Jiang}, \bibinfo{author}{P.~Hu}, \bibinfo{author}{Y.~Li}, \bibinfo{author}{C.~Yuan}, \bibinfo{author}{I.~Masood}, \bibinfo{author}{H.~Jelodar}, \bibinfo{author}{M.~Rabbani}, \bibinfo{author}{Y.~Wang},
\newblock \bibinfo{title}{A survey of real-time approximate nearest neighbor query over streaming data for fog computing},
\newblock \bibinfo{journal}{Journal of Parallel and Distributed Computing} \bibinfo{volume}{116} (\bibinfo{year}{2018}) \bibinfo{pages}{50--62}.
\bibitem[{Johnson et~al.(2019)Johnson, Douze, and J{\'e}gou}]{johnson2019billion}
\bibinfo{author}{J.~Johnson}, \bibinfo{author}{M.~Douze}, \bibinfo{author}{H.~J{\'e}gou},
\newblock \bibinfo{title}{Billion-scale similarity search with gpus},
\newblock \bibinfo{journal}{IEEE Transactions on Big Data} \bibinfo{volume}{7} (\bibinfo{year}{2019}) \bibinfo{pages}{535--547}.
\bibitem[{Zong et~al.(2023)Zong, Song, and Liu}]{zong2023detrs}
\bibinfo{author}{Z.~Zong}, \bibinfo{author}{G.~Song}, \bibinfo{author}{Y.~Liu},
\newblock \bibinfo{title}{Detrs with collaborative hybrid assignments training},
\newblock in: \bibinfo{booktitle}{Proceedings of the IEEE/CVF international conference on computer vision}, \bibinfo{year}{2023}, pp. \bibinfo{pages}{6748--6758}.
\bibitem[{Lin et~al.(2014)Lin, Maire, Belongie, Hays, Perona, Ramanan, Doll{\'a}r, and Zitnick}]{lin2014microsoft}
\bibinfo{author}{T.-Y. Lin}, \bibinfo{author}{M.~Maire}, \bibinfo{author}{S.~Belongie}, \bibinfo{author}{J.~Hays}, \bibinfo{author}{P.~Perona}, \bibinfo{author}{D.~Ramanan}, \bibinfo{author}{P.~Doll{\'a}r}, \bibinfo{author}{C.~L. Zitnick},
\newblock \bibinfo{title}{Microsoft coco: Common objects in context},
\newblock in: \bibinfo{booktitle}{Computer Vision--ECCV 2014: 13th European Conference, Zurich, Switzerland, September 6-12, 2014, Proceedings, Part V 13}, \bibinfo{organization}{Springer}, \bibinfo{year}{2014}, pp. \bibinfo{pages}{740--755}.
\bibitem[{Deng et~al.(2009)Deng, Dong, Socher, Li, Li, and Fei-Fei}]{deng2009imagenet}
\bibinfo{author}{J.~Deng}, \bibinfo{author}{W.~Dong}, \bibinfo{author}{R.~Socher}, \bibinfo{author}{L.-J. Li}, \bibinfo{author}{K.~Li}, \bibinfo{author}{L.~Fei-Fei},
\newblock \bibinfo{title}{Imagenet: A large-scale hierarchical image database},
\newblock in: \bibinfo{booktitle}{2009 IEEE conference on computer vision and pattern recognition}, \bibinfo{organization}{Ieee}, \bibinfo{year}{2009}, pp. \bibinfo{pages}{248--255}.
\bibitem[{He et~al.(2016)He, Zhang, Ren, and Sun}]{he2016deep}
\bibinfo{author}{K.~He}, \bibinfo{author}{X.~Zhang}, \bibinfo{author}{S.~Ren}, \bibinfo{author}{J.~Sun},
\newblock \bibinfo{title}{Deep residual learning for image recognition},
\newblock in: \bibinfo{booktitle}{Proceedings of the IEEE conference on computer vision and pattern recognition}, \bibinfo{year}{2016}, pp. \bibinfo{pages}{770--778}.
\bibitem[{Nguyen et~al.(2024)Nguyen, Tran, Pham, and Jeon}]{nguyen2024optimizing}
\bibinfo{author}{H.-H. Nguyen}, \bibinfo{author}{D.~N.-N. Tran}, \bibinfo{author}{L.~H. Pham}, \bibinfo{author}{J.~W. Jeon},
\newblock \bibinfo{title}{Optimizing monocular driving assistance for real-time processing on jetson agx xavier},
\newblock \bibinfo{journal}{IEEE Access}  (\bibinfo{year}{2024}).
\bibitem[{Pace et~al.(2018)Pace, Aloi, Gravina, Caliciuri, Fortino, and Liotta}]{pace2018edge}
\bibinfo{author}{P.~Pace}, \bibinfo{author}{G.~Aloi}, \bibinfo{author}{R.~Gravina}, \bibinfo{author}{G.~Caliciuri}, \bibinfo{author}{G.~Fortino}, \bibinfo{author}{A.~Liotta},
\newblock \bibinfo{title}{An edge-based architecture to support efficient applications for healthcare industry 4.0},
\newblock \bibinfo{journal}{IEEE Transactions on Industrial Informatics} \bibinfo{volume}{15} (\bibinfo{year}{2018}) \bibinfo{pages}{481--489}.
\bibitem[{Kavalionak et~al.(2019)Kavalionak, Gennaro, Amato, Vairo, Perciante, Meghini, and Falchi}]{kavalionak2019distributed}
\bibinfo{author}{H.~Kavalionak}, \bibinfo{author}{C.~Gennaro}, \bibinfo{author}{G.~Amato}, \bibinfo{author}{C.~Vairo}, \bibinfo{author}{C.~Perciante}, \bibinfo{author}{C.~Meghini}, \bibinfo{author}{F.~Falchi},
\newblock \bibinfo{title}{Distributed video surveillance using smart cameras},
\newblock \bibinfo{journal}{Journal of Grid Computing} \bibinfo{volume}{17} (\bibinfo{year}{2019}) \bibinfo{pages}{59--77}.

\end{thebibliography}
